\documentstyle[11pt,newpasp,twoside,epsf]{article}
\def\edcomment#1{\iffalse\marginpar{\raggedright\sl#1\/}\else\relax\fi}
\reversemarginpar

\begin{document}
\title{Formation of globular clusters in galaxy mergers}
\author{Kenji Bekki, Warrick J. Couch}
\affil{School of physics, University of New Southwales, Sydney, NSW, 2052, Australia
}
\author{Duncan A. Forbes and M. A. Beasley}
\affil{Centre for Astrophysics \& Supercomputing, Swinburne University
of Technology, Hawthorn, VIC,  3122, Australia}

\begin{abstract}
Our numerical simulations  first demonstrate that the pressure of ISM in a major merger 
becomes so high  ($>$  $10^5$ $\rm k_{\rm B}$ K $\rm cm^{-3}$) that
GMCs in the merger can collapse to form globular clusters (GCs) within a few Myr.
The star formation efficiency within a GMC in galaxy mergers can rise up
from a few percent to $\sim$ 80 percent, depending on the shapes and the
temperature of the GMC. This implosive GC formation due to external
high pressure of warm/hot ISM can be more efficient  in
the tidal tails or the central regions of mergers. 
The developed clusters
have King-like profile with the effective radius of a few pc.  
The structural, kinematical, and chemical properties of these GC systems
can depend on orbital and chemical properties of major mergers. 
\end{abstract}
\section{Implosive formation of globular clusters}

Several authors have suggested that very high pressure of ISM expected
in major galaxy merging can be responsible for the rapid (triggered) collapse of GMCs and
the subsequent GC formation (e.g., Elmegreen \& Efremov 1997; Bekki et al. 2002).
We numerically investigate the key questions on this GC formation scenario: whether or not
the star formation efficiency within a GMC under such high pressure of ISM
can be as high as that ($>$ 50 \%, e.g., Hills 1980) required for the bound 
cluster formation. 
Figure 1 describes that 
the high pressure of the ISM can continue to strongly compress 
the cloud without losing a significant amount of gas from the cloud.
As the strong compression proceeds, 
the internal density/pressure of the cloud can rise so significantly
that a GC can form from the gas with a starburst 
(See the caption of Fig.1 for the detail of this).

\begin{figure}
\plotone{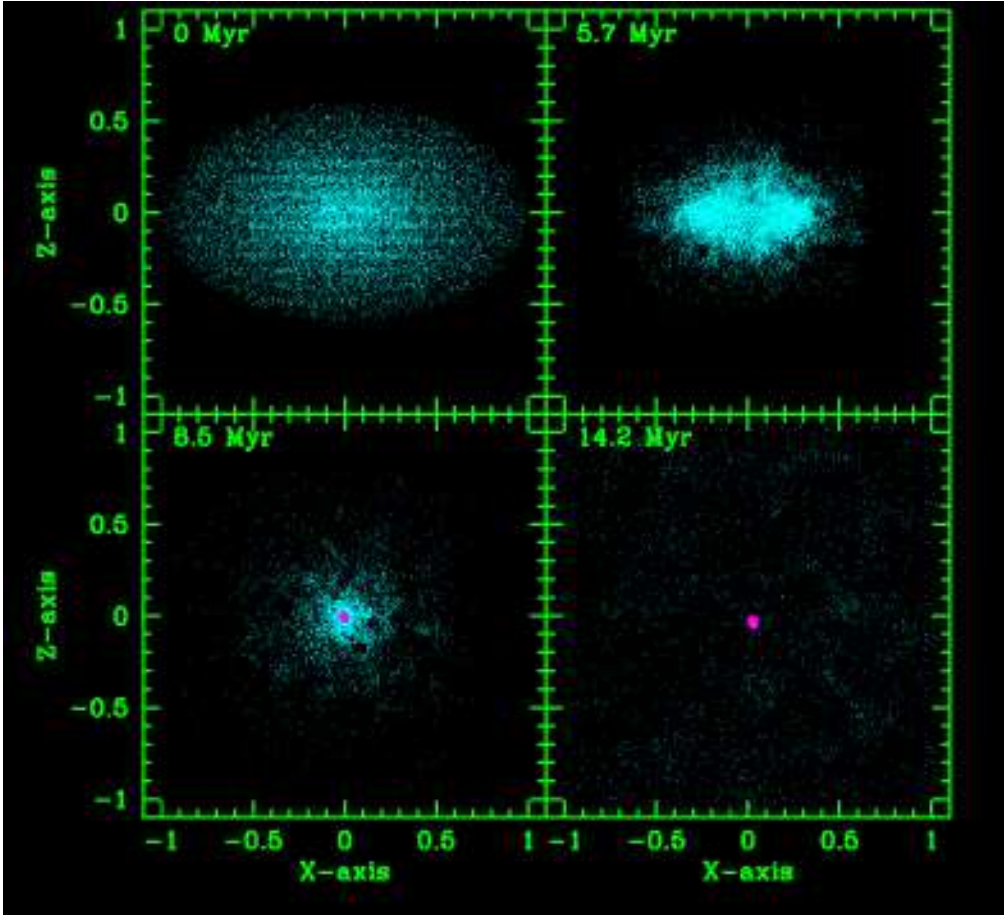}
\caption{Time evolution of a GMC embedded by high pressure ISM for gaseous
components (cyan) and for new stars formed from the gas (magenta).
For clarity, the surrounding high pressure warm/hot ISM
with $P$ $\sim$  $10^5$ $\rm k_{\rm B}$ K $\rm cm^{-3}$ 
is not shown.
The mass and the size of the initial GMC is $ 10^6 \rm M_{\odot}$ and 97 pc,
respectively. The oblate shape of the cloud is assumed
in this model (the long-to-short-axis-ratio is set to be 0.6).
The initial sound velocity of the gas in GMC is $\sim$ 5 km s$^{-1}$,
for which the GMC can not collapse owing to its self-gravity 
(if the surrounding ISM's density and pressure are  as low as  those
observed in disk galaxies). 
One frame measures 200 pc on a side. Note that a very compact GC 
is formed owing to strong external compression of the GMC by
the hot, high pressure ISM. 
About 80 \% of the initial gas is converted into
new stars within a few Myr so that the developed compact stellar system 
with the effective radius of a few  pc
can be strongly bounded even after the removal of the remaining gas.
Due to the rapid, dissipative collapse, 
the gaseous density of the cloud dramatically rises and 
consequently star formation begins in the central regions of the cloud.
The star formation rate increase significantly 
to 1.5 $M_{\odot}$ yr$^{-1}$ (8\,Myr after the start of the cloud's collapse).
Because of the ``implosive'' formation of stars from
strongly compressed gas, the developed stellar system is strongly
self-gravitating and compact. This result implies that high external pressure
from the ISM is likely to trigger the formation of bound, compact star
clusters rather than unbound, diffuse field stars. 
}
\end{figure}

\end{document}